\documentstyle[prl,aps,epsfig,floats]{revtex}
\begin{document}
\draft

\def\lsi{\raise0.3ex\hbox{$<$\kern-0.75em\raise-1.1ex\hbox{$\sim$}}}
\def\gsi{\raise0.3ex\hbox{$>$\kern-0.75em\raise-1.1ex\hbox{$\sim$}}}
\newcommand{\lsim}{\mathop{\lsi}}
\newcommand{\gsim}{\mathop{\gsi}}
\def\hxi{\hat{\xi}}
\def\hk{\hat{k}}
\def\bE{\mbox{\boldmath ${E}$}}
\def\bB{\mbox{\boldmath ${B}$}}
\def\bD{\mbox{\boldmath ${D}$}}
\def\vA{\vec A}
\def\vE{\vec E}
\def\vB{\vec B}
\def\vD{\vec D}
\def\vn{\vec\nabla}
\def\cH{{\mathcal H}}
\def\half{\frac{1}{2}}
\textheight=23.8cm

\twocolumn[\hsize\textwidth\columnwidth\hsize\csname@twocolumnfalse\endcsname

\title{Defect formation and local gauge invariance}
\author{M.~Hindmarsh
and A.~Rajantie}
\address{Centre for Theoretical Physics,
University of Sussex, Falmer, Brighton BN1 9QJ,~~~U.~K. }

\date{\today}
\maketitle
\begin{abstract}
We propose a new mechanism for formation of topological defects in a U(1)
model with a local gauge symmetry. 
This mechanism leads to definite predictions,
which are qualitatively different from those of the Kibble-Zurek mechanism
of global theories. We confirm these predictions in numerical simulations,
and they can also be tested in superconductor experiments.
We believe that the mechanism generalizes to more complicated theories.
\end{abstract}
\pacs{PACS numbers: 11.15.Ex, 11.27.+d, 74.60.Ge
\hfill SUSX-TH-00-012
\hfill hep-ph/0007361 
}
\vskip.5pc
]
%%%%%%%%%%%%%%%%%%%%%%%%%%%%%%%%%%%%%%%%%%%%%%%%%%%%%%

When a global symmetry is spontaneously
broken in a phase transition, 
it is generally accepted that the formation of topological
defects is well described by the Kibble-Zurek (KZ) 
scenario~\cite{Kibble:1976sj,Zurek:1985qw}.
As the transition takes place in a finite time,
the
correlation length of the order parameter cannot keep up with its
equilibrium value, which diverges at the transition point. 
The maximum correlation length reached determines
the average distance of the defects in the final 
state~\cite{Kibble:1976sj}, and
it can be estimated from the critical dynamics of the 
theory~\cite{Zurek:1985qw}.
This scenario has been tested in many numerical 
simulations~\cite{Laguna:1997pv,Laguna:1998cf}
and in experiments with 
$^3$He~\cite{Ruutu:1996qz}
 and $^4$He~\cite{ref:He4}. Although the experimental
results for $^4$He are in disagreement 
with the theory, the general picture
is believed to be correct.

However, in theories where the relevant symmetry is a local gauge
invariance,
e.g.~in superconductors or in cosmology~\cite{Hindmarsh:1995re}, 
the validity of the
KZ scenario has been questioned~\cite{Kibble:1995aa}, although
lattice 
simulations~\cite{Yates:1998kx} have found compatible results.
Evidence that the KZ scenario might not work in
these cases has been provided by recent experiments on
YBCO superconductors~\cite{ref:carmi}. 
Instead of detecting individual vortices, the authors measured the total
net flux through the whole system, and found it to be zero within
the accuracy of the experimental setup, in contradiction with predictions
of the KZ scenario.

The purpose of this letter is to suggest 
a simple and intuitive picture
for defect formation in theories
with local gauge symmetries.
This picture is quite different from the KZ scenario
and leads to definite predictions, which we have confirmed in
numerical simulations. Furthermore, they can also be tested in
superconductor experiments.

Let us first review the KZ picture for 
the global case and consider for simplicity a U(1) 
symmetric field theory in $D$ spatial
dimensions. In the broken
phase, the vacuum manifold is topologically a circle and 
the topological defects are therefore vortices with dimensionality
$D-2$. The Hamiltonian contains a gradient
term $|\vn\phi|^2$, where $\phi=v\exp(i\theta)$ 
is the order parameter field. 
Deep in the broken phase, the gradient
term implies that $\vn\theta\approx 0$ in equilibrium.
Thus the phase angle is correlated at infinitely
long distances, but since it takes an infinitely long time for
the system to achieve that, 
after the transition
$\theta$ will be approximately constant only inside domains of 
size $\hxi$, given by the maximum correlation length reached
during the transition,
and its value will be uncorrelated between these 
domains~\cite{Kibble:1976sj}.
Since vortices are characterized by a non-zero change of $\theta$ around
a closed loop, they are formed
where three domains meet with a probability that is independent of 
the size of the domains.
Consequently, 
the final vortex number in the global case
behaves as $N\sim \hxi^{-2}$.
The value of $\hxi$ can be estimated from the critical slowing
down of the dynamics during the 
transition~\cite{Zurek:1985qw}.

If the symmetry is local, there is another, competing mechanism,
which dominates if the transition is sufficiently slow. 
%
% NEW STUFF 000910
%
In the temporal gauge $A_0=0$,
the Hamiltonian for a relativistic gauged U(1) scalar field theory 
in three spatial dimensions is
\begin{equation}
H = \int d^3x \left[
\half {\vE}^2 + \half {\vB}^2 + |\Pi|^2 + |\vD\phi|^2 + V(\phi)
\right],
\label{e:Ham}
\end{equation}
where $\Pi=\partial_0\phi$ 
is the canonical momentum, $\vE=-\partial_0\vA$ and $\vB=
\vec{\nabla}\times\vA$ the electric and
magnetic field strengths, and $V(\phi)$ the potential of the scalar
field.
The corresponding equations of motion are
\begin{eqnarray}
\partial^2_0\phi &=& \vD^2\phi-V'(\phi),
\nonumber\\
\partial_0 \vE&=&\vn\times\vB+2e{\rm Im}\phi^*\vD\phi,
\nonumber\\
\vn\cdot\vE&=&2e{\rm Im}\phi^*\partial_0\phi.
\label{equ:eom}
\end{eqnarray}
(Note that we use units with $k_B=\hbar=c=\mu_0=1$.)
More generally, we will consider the analogous system in $D$ spatial 
dimensions.
%
% END NEW STUFF
%
Because the gradient has been replaced by a
covariant derivative
$\vD\phi=\vn\phi+ie\vA\phi$,
the energy is minimized in the broken phase
if
$\vn\theta\approx -e\vA$. 
In the presence of a magnetic flux, this condition cannot be satisfied
everywhere and frustrations, vortices, are formed.
Although the
magnetic flux is zero on the average, 
the thermal fluctuations give it a non-zero variance.
When the system enters the broken phase, it
tries to rearrange the field configuration
to minimize the magnetic flux
and the energy associated with it. Because of
the finite time available, this is not possible for the fluctuations with
the longest wavelengths
and they freeze in their initial form, but at shorter distances
the fluctuations of the magnetic flux are smoothed out.
Therefore, immediately after the transition, before the flux is
localized into vortices, its configuration consists of domains inside which 
it is approximately uniform and which have some characteristic 
size $\hxi$.

If we calculate the winding
number around a curve $C$, which encircles
one of the domains discussed above,
it typically does not vanish, but 
instead
\begin{eqnarray}
\label{equ:ncphi}
n_C&\equiv& \frac{1}{2\pi}\int_C d\vec{r}\cdot\vec{\nabla}\theta
\approx -\frac{e}{2\pi} \int_C d\vec{r}\cdot\vec{A}
\equiv 
-\frac{e}{2\pi}\Phi_C,
\end{eqnarray}
where $\Phi_C$ is the magnetic flux through
the curve $C$. 
Because the flux rearrangements at the transition were
only able to change its distribution inside the domain, $\Phi_C$ has the
same value it had in the symmetric phase. We can estimate it
in the
standard way by calculating the energy $E(\Phi_C)$ associated with
the flux. Inside the domains, the flux is uniform and therefore
\begin{equation}
E(\Phi_C)\approx\hxi^D\left(\frac{\Phi_C}{\hxi^2}\right)^2
=\hxi^{D-4}\Phi_C^2.
\end{equation}
Requiring
$E(\Phi_C)\approx T$ 
gives the typical value of the flux
\begin{equation}
\label{equ:flsize}
\Phi_C\approx T^{1/2}\hxi^{2-D/2}.
\end{equation}

Using Eqs.~(\ref{equ:ncphi}) and (\ref{equ:flsize}), we can 
now estimate the area density of vortices after the transition.
Suppose we have in our $D$-dimensional space a surface of area $A$,
then it is split into $\sim A\hxi^{-2}$ domains in the transition. Each
domain
contains $N_0\approx(e/2\pi)\Phi_C$ vortices, and we will
assume that $N_0\gsim 1$.
Then, the total number of
vortices piercing the surface per unit area is
\begin{equation}
\label{equ:power2d}
N/A\approx \frac{e}{2\pi}T^{1/2}\hxi^{-D/2}.
\end{equation}
In particular, at $D=2$, $N\sim\hxi^{-1}$.

In addition, vortices will also be formed by a variant of the 
KZ mechanism. Initially, these vortices behave like global vortices,
but eventually a quantum of magnetic flux is generated inside
each of them, making them truly local vortices.
However, the KZ mechanism is only important if
the transition is very rapid or the temperature very low, i.e.~$N_0\lsim 1$.
Thus we will neglect it in the following. 

At $D=3$, the vortices formed inside the domains form a 
network at distances longer than
$\hxi$, and in this network most vortices will be in the form of closed
loops, which will quickly shrink into a point and disappear. Therefore it
is useful to consider a border-line case between $D=2$ and $D=3$,
where one of the dimensions $L_z$ is very short. 
As long as $L_z\lsim \hxi$,
the vortices will wind around the short dimension rather than
forming loops and therefore they will be stable.
Now the domains have the form of a short cylinder, and the estimate
(\ref{equ:power2d}) generalizes easily to this case, yielding
\begin{equation}
\label{equ:N0cyl}
N/A\approx
\frac{e}{2\pi}T^{1/2}L_z^{-1/2}\hxi^{-1}
\sim L_z^{-1/2}.
\end{equation}

The
correlations between the vortices in the final state will 
also be different from
those predicted by the KZ mechanism. 
Let us consider a system after the transition and
assume that there is a vortex with a positive winding
at point $\vec{x}$. In the KZ scenario,
the distance to the nearest other vortex should be roughly $\hxi$.
If we calculate
the winding number $n_C(r)$ of a circular loop $C$ of radius $r$
centered at $\vec{x}$, it follows
that at $r\lsim\hxi$, the winding number is close to one. However,
at distances $r\gsim\hxi$, the phase angle is independent of whether
there is a vortex inside at $\vec{x}$ or not, and therefore
$n_C(r)=0$. Thus, in the KZ scenario,
\begin{equation}
\label{equ:globNc}
n_C(r)\approx\left\{
\begin{array}{ll}
1,&r\lsim\hxi,\\
0,&r\gsim\hxi.
\end{array}
\right.
\end{equation}

Again, our case is very different. Inside a single domain,
all vortices have the same sign, and $n_C(r)$ is therefore an
increasing function of $r$ at $r\lsim\hxi$. At $r\gsim\hxi$, $n_C(r)$ 
gets
contributions
from different domains, and since they are all independent they average to
zero and $n_C(r)$ becomes 
a constant. The behaviour in our scenario is therefore
totally
different from Eq.~(\ref{equ:globNc}):
\begin{equation}
\label{equ:locNc}
n_C(r)\approx\left\{
\begin{array}{ll}
1+c_0r^2,&r\lsim\hxi,\\
c_1,&r\gsim\hxi,
\end{array}
\right.
\end{equation}
where $c_0\ge 0$ and $c_1\ge 1$ are constants.

Finally, let us discuss the dependence of $\hxi$ on the 
``quench'' timescale $\tau_Q$,
which parameterizes the rate at which the phase transition takes place.
For definiteness, we consider the potential
\begin{equation}
V(\phi)=m^2(t)|\phi|^2+\lambda|\phi|^4,
\end{equation}
where the mass parameter is changed linearly across its critical value
\begin{equation}
\label{equ:mtideal}
m(t)^2=m_c^2-\delta m^2\frac{t}{\tau_Q}.
\end{equation}
In reality, $m_c^2$ is not equal to zero when thermal
fluctuations are taken into account, but let us use the mean-field
approximation in the following, in which case it is.

If we know the photon dispersion relation $\omega=\omega(k)$ at the
transition point and its neighbourhood, we can estimate that
a Fourier mode of wave number $k_i$ falls out of equilibrium
during the transition if the adiabaticity condition is not satisfied
\begin{equation}
\label{equ:adiab}
\left|\frac{d\omega(k)}{dt}\right|>|\omega(k)|^2.
\end{equation}
However, the calculation of the dispersion relation is beyond the
scope of this letter, and
we will instead only consider two simple special
cases: the overdamped case (OD), where the dynamics is dominated
by a $k$-independent damping rate $\gamma$, 
and the underdamped case (UD) with the free-field
dispersion relation:
\begin{equation}
\label{equ:disp}
\omega=
i\gamma^{-1}(k^2+m_\gamma^2)~{\rm(OD)},\quad
(k^2+m_\gamma^2)^{1/2}~{\rm(UD)}.
\end{equation}
These same special cases have been discussed in the context of global theories
in Ref.~\cite{Laguna:1998cf}, but we stress that it is by no means clear that
the dynamics is well described by either of these cases. 
Furthermore, we assume that the photon mass for the relevant modes 
behaves as
$m_\gamma^2\approx 2e^2 |\phi|^2$, and that $|\phi|^2$ is given by 
its equilibrium
value $|\phi|^2=-m^2(t)/2\lambda\sim t/\tau_Q$. 

Now, Eq.~(\ref{equ:adiab})
tells us that the highest wave numbers that fall out of equilibrium
behave as 
\begin{equation}
\label{equ:noneqk}
\hk\sim
\tau_Q^{-1/4}~{\rm(OD)},\quad\tau_Q^{-1/3}~{\rm(UD)}.
\end{equation}
The domain size $\hxi$ is then simply given by $\hxi\approx 2\pi/\hk$,
and using Eq.~(\ref{equ:power2d}), we can write down the dependence of the
final vortex number on $\tau_Q$ in two dimensions as
\begin{equation}
\label{equ:localexp}
N\sim\tau_Q^{-1/4}~{\rm(OD)},\quad\tau_Q^{-1/3}~{\rm(UD)}.
\end{equation}
In the KZ scenario, the analogous exponents 
are $-1/2$ and $-2/3$, respectively~\cite{Laguna:1998cf}.

We carried out a set of numerical simulations 
%
% NEW STUFF 000910
%
using the equations of motion 
(\ref{equ:eom}) % in the temporal gauge $A_0=0$ 
%
% END NEW STUFF
%
to test
the results in Eqs.~(\ref{equ:N0cyl}), (\ref{equ:locNc})
and (\ref{equ:localexp}).
Our coupling constants were $e=0.3$ and $\lambda=0.18$. Since $\lambda>e^2$,
the transition is continuous, as in a Type II superconductor.
We used periodic lattices
of size $120\times 120\times L_z$, where $L_z=5$ or 20
(in units where the lattice spacing is one).
We prepared a set of initial conditions according to the thermal
distribution $\exp(-H/T)$ at $T=6$ using
%After thermalizing the system to temperature $T=6$ using 
a hybrid
Monte Carlo algorithm and
followed the time evolution using a leap-frog algorithm with time step
$\delta t=0.05$, 
changing the mass parameter according to
\begin{equation}
m^2(t)=m_0^2-\delta m^2\left(\frac{4}{3\pi}\arctan\frac{t}{\tau_Q}+\frac{1}{3}
\right).
\label{equ:masschange}
\end{equation}
Here $m_0^2=-1.6$ is the initial value at 
$t=-\tau_Q$ when the quench begins and
$\delta m^2=3.2$ was chosen
in such a way that the transition takes place at $t\approx 0$.
As in Eq.~(\ref{equ:mtideal}), the rate of change of
$m^2$ is proportional to $\tau_Q$, but
in this form, we reach an equilibrium state at late times,
which makes comparison of different quench rates easier.
The details of the simulations will be discussed in Ref.~\cite{ref:htlvort}.

\begin{figure}
\epsfig{file=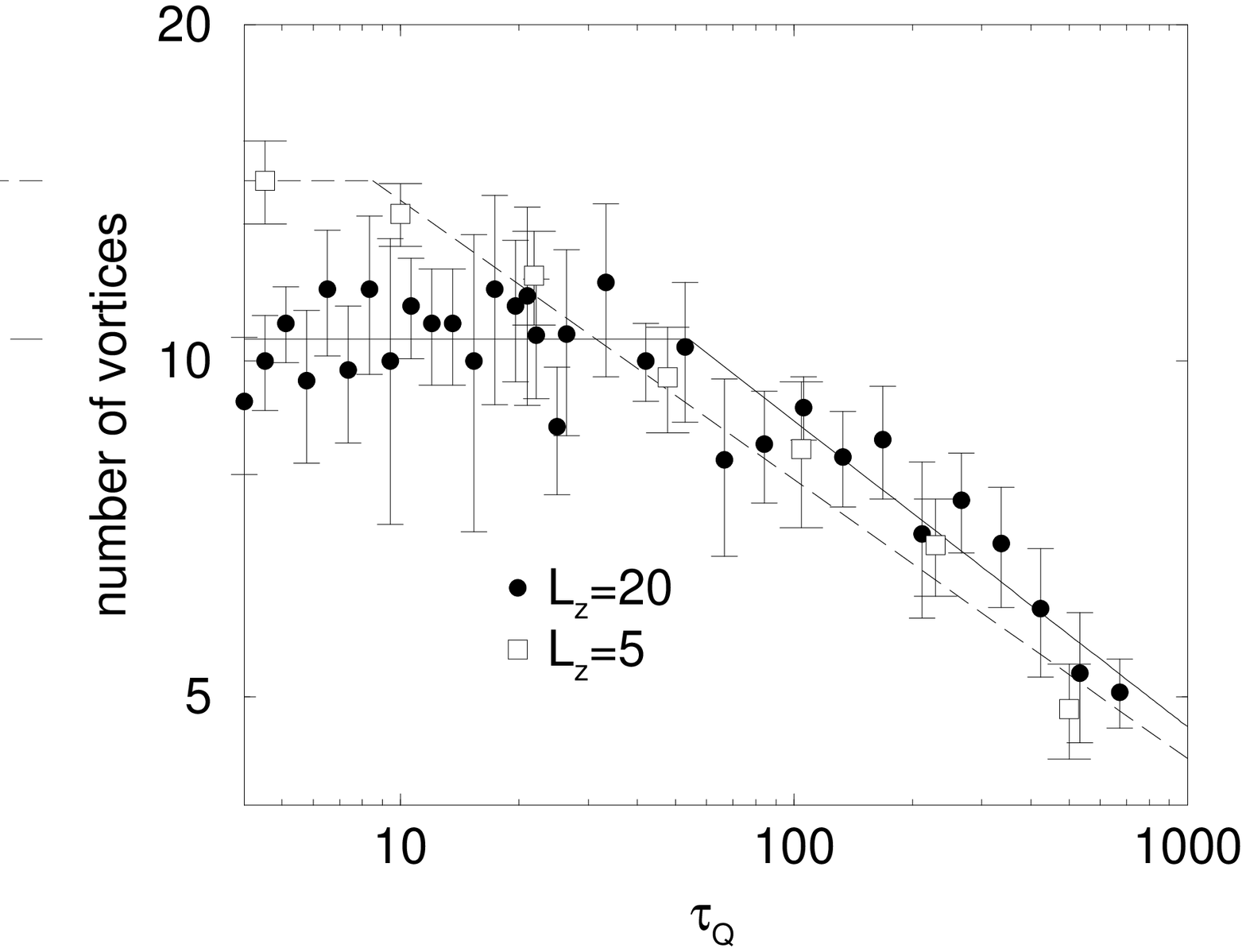,width=7.5cm}
\caption{The number of vortices in the final state as a function of $\tau_Q$.
The $L_z=5$ data have been scaled by a factor of $1/2$ in
accordance with Eq.~(\ref{equ:N0cyl}).
The solid and dashed lines are the fits to Eq.~(\ref{equ:fit}) (see Table
\ref{table:fit})
and correspond to the cases $L_z=20$ and $L_z=5$, respectively.}
\label{fig:taudep}
\end{figure}

In the final state, the vortices have only short-range interactions
and therefore they freeze quickly into their final configuration.
After that,
at $t=\tau_Q+400$, we located the vortices from the field 
configuration by measuring the gauge-invariant winding numbers of
individual plaquettes~\cite{Kajantie:1998bg} 
and connecting plaquettes with
non-zero winding numbers into vortex lines.
We then measured the total number of those vortex lines that wind around
the short dimension of our lattice. The results are shown in 
Fig.~\ref{fig:taudep} as a function of $\tau_Q$. The results from the
thinner lattice $L_z=5$ have been divided by 2, because according 
to Eq.~(\ref{equ:N0cyl}) the results
from the two lattices
should then be on top of each other.
Each datapoint is an average of $\sim 20$ runs starting from different
initial configurations drawn from the thermal ensemble.

In slow transitions, the data agree very well with the
predicted power-law behaviour.
At $\tau_Q\lsim 40-50$, the $L_z=20$ data become independent of
$\tau_Q$, which suggests that $\hxi<L_z$, and part of the vortices
have formed loops and do not contribute to the vortex count.
We fitted the function 
\begin{equation}
\label{equ:fit}
f(\tau_Q;c,\tau_c,\alpha)=\left\{
\begin{array}{ll}
c\tau_c^{-\alpha},&\tau_Q<\tau_c\\
c\tau_Q^{-\alpha},&\tau_Q>\tau_c
\end{array}
\right.
\end{equation}
into the data, and the results are
given in Table~\ref{table:fit}.
Both cases agree with the overdamped prediction $\alpha=-0.25$.
Moreover, the ratio of the results from $L_z=5$ and $L_z=20$ measured
at, say, $\tau_Q=100$, is
$1.8\pm0.2$, which is compatible with the value
2 predicted by Eq.~(\ref{equ:N0cyl}).

Using our data, we 
can also perform a more quantitative check for Eq.~(\ref{equ:N0cyl})
even without knowledge of the
dynamics that determines $\hxi$,
if we assume that $\hxi=L_z$ at the turning point 
in the $L_z=20$ data. By substituting this into
Eq.~(\ref{equ:N0cyl}), we find that the number of vortices
formed at $\tau_Q<\tau_c$ in the $L_z=20$ case should be
$N\approx 19$,
which agrees within a factor of 2 with the numerical result.

\begin{table}
$$
\begin{array}{c|c|c|c|c}
& c & \tau_c & \alpha & \chi^2 \\
\hline
L_z=5 & 49.5\pm 2.6 & 8.5 \pm 1.7 & 0.250\pm 0.013 & 6.73\\
L_z=20 & 31.2\pm 6.7 & 54.1 \pm 13.7 & 0.274\pm 0.039 & 12.09
\end{array}
$$
\caption{Fit of the data to the functional form (\ref{equ:fit}).
Results from both simulations are compatible with the overdamped
exponent $\alpha=0.25$ (see Eq.~(\ref{equ:localexp})).}
\label{table:fit}
\end{table}

\begin{figure}
\epsfig{file=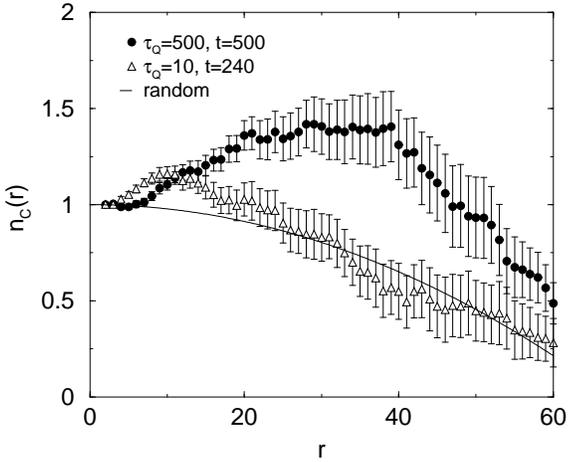,width=7.5cm}
\caption{The net number of vortices within distance $r$ from a vortex
with positive winding in a slow (circles) and a fast (triangles)
quench on a thin lattice ($L_z=5$). 
The dotted line is the corresponding curve
for uncorrelated vortex-antivortex pairs.
The fact that the data points
are well above the random curve shows that there is a positive correlation 
between vortices at short distances.}
\label{fig:Nc}
\end{figure}

We can also test Eq.~(\ref{equ:locNc}) by measuring the net number of vortices
inside a circle of radius $r$ centered at a vortex. 
The results for
two different values of $\tau_Q$ with $L_z=5$ are shown in
Fig.~\ref{fig:Nc}, together with the benchmark curve $n_C=1-\pi r^2A^{-1}$,
which corresponds to
a uniform random distribution of vortex-antivortex pairs
and shows the effect of the finite system size.
At short distances,
the data points are significantly
above this curve, which indicates a positive correlation
between vortices, and
at long distances, they follow the benchmark curve.
Both of these results agree with Eq.~(\ref{equ:locNc}) but differ
from the KZ prediction~(\ref{equ:globNc}).

Our results disagree with the simulations 
in Ref.~\cite{Yates:1998kx}, where the authors found
an exponent $\alpha$ that was compatible with the global theory.
We believe that the temperature used there, 
$T=0.01$, 
was so low that
practically all of the vortices were formed by the KZ mechanism.

Since our mechanism can only increase the number of vortices
formed in a transition from that predicted by the KZ mechanism, it 
cannot explain the
failure in Ref.~\cite{ref:carmi} to find any total net magnetic flux
when a superconductor film was quenched into the superconducting phase.
However, the experiment does not rule it out either, because
the vortices formed by the KZ mechanism might avoid detection
by being expelled from the film before generating observable
magnetic flux.
The extra magnetic flux predicted 
by our mechanism is fairly 
small, because it can only change the flux distribution at short distances.
The minimal energy
for a configuration with a
given value $\Phi$ of flux through the film is that of a 
magnetic dipole, 
$E_{\rm min}\approx\mu_0^{-1}A^{-1/2}\Phi^2$.
Using the values $T=90$~K and $A=1$~cm$^2$, 
we find that the predicted number of flux quanta is
\begin{equation}
N\approx \frac{e}{\pi\hbar}(\mu_0k_BT)^{1/2}A^{1/4}
\approx 2,
\end{equation}
which is below the resolution of the experiment.
A similar estimate applies for a recent Josephson junction 
experiment~\cite{ref:carmi2},
where $N\approx 7$ flux quanta were observed. In this case, the
prediction of the KZ scenario is $N\approx 4$, and thus the experiment
cannot decide between the two mechanisms.
It seems that our scenario can only be confirmed by experiments in which
not only the total flux but also the spatial distribution of the vortices
can be measured.

To summarize, our numerical simulations confirm
the results in Eqs.~(\ref{equ:N0cyl}) and (\ref{equ:locNc}), 
which are independent of
assumptions about the dispersion relation $\omega(k)$.
Using the naive overdamped dispersion relation (\ref{equ:disp}),
we can also reproduce accurately
the exponents $\alpha$ in Table~\ref{table:fit}.
This supports strongly the scenario
presented in this letter.

%In order to make more quantitative predictions, it is crucial to
%understand the dispersion relation $\omega(k)$ in the neighbourhood
%of the transition point, and we will discuss that in a future 
%publication~\cite{ref:htlvort}. We also believe that the mechanism presented
%here can easily be generalized to other types of gauged topological defects.

%\acknowledgements
AR was supported by PPARC and also partly by the University of Helsinki.
Part of this work was conducted on the SGI Origin platform using COSMOS
Consortium facilities, funded by HEFCE, PPARC and SGI.
We acknowledge computing support from
the Sussex High Performance Computing Initiative.

\end{document}